\documentclass[twocolumn,amsmath,amssymb,10pt]{revtex4}
\reversemarginpar
\usepackage{labelfig,graphicx}

\usepackage{graphicx}
\usepackage{dcolumn}
\usepackage{bm}
\usepackage{amsfonts}
\usepackage{latexsym}

\begin{document}
\newcommand{\commentout}[1]{}

\newcommand{\nwc}{\newcommand}
\newcommand{\bz}{{\mathbf z}}
\newcommand{\sqk}{\sqrt{\ks}}
\newcommand{\sqkone}{\sqrt{|\ks_1|}}
\newcommand{\sqktwo}{\sqrt{|\ks_2|}}
\newcommand{\invsqkone}{|\ks_1|^{-1/2}}
\newcommand{\invsqktwo}{|\ks_2|^{-1/2}}
\newcommand{\partz}{\frac{\partial}{\partial z}}
\newcommand{\grady}{\nabla_{\by}}
\newcommand{\gradp}{\nabla_{\bp}}
\newcommand{\gradx}{\nabla_{\bx}}
\newcommand{\invf}{\cF^{-1}_2}
\newcommand{\myphi}{\Phi_{(\eta,\rho)}}
\newcommand{\minrg}{|\min{(\rho,\gamma^{-1})}|}
\newcommand{\al}{\alpha}
\newcommand{\xvec}{\vec{\mathbf x}}
\newcommand{\kvec}{{\vec{\mathbf k}}}
\newcommand{\lt}{\left}
\newcommand{\ksq}{\sqrt{\ks}}
\newcommand{\rt}{\right}
\newcommand{\ga}{\gamma}
\newcommand{\vas}{\varepsilon}
\newcommand{\lan}{\left\langle}
\newcommand{\ran}{\right\rangle}
\newcommand{\tvas}{{W_z^\vas}}
\newcommand{\psiep}{{W_z^\vas}}
\newcommand{\wep}{{W^\vas}}
\newcommand{\weptil}{{\tilde{W}^\vas}}
\newcommand{\wepz}{{W_z^\vas}}
\newcommand{\weps}{{W_s^\ep}}
\newcommand{\wepsp}{{W_s^{\ep'}}}
\newcommand{\wepzp}{{W_z^{\vas'}}}
\newcommand{\wepztil}{{\tilde{W}_z^\vas}}
\newcommand{\vvas}{{\tilde{\ml L}_z^\vas}}
\newcommand{\veptil}{{\tilde{\ml L}_z^\vas}}
\newcommand{\vep}{{{ V}_z^\vas}}
\newcommand{\cvc}{{{\ml L}^{\ep*}_z}}
\newcommand{\cvcp}{{{\ml L}^{\ep*'}_z}}
\newcommand{\cvp}{{{\ml L}^{\ep*'}_z}}
\newcommand{\cvtil}{{\tilde{\ml L}^{\ep*}_z}}
\newcommand{\cvtilp}{{\tilde{\ml L}^{\ep*'}_z}}
\newcommand{\vtil}{{\tilde{V}^\ep_z}}
\newcommand{\ktil}{\tilde{K}}
\newcommand{\n}{\nabla}
\newcommand{\tkappa}{\tilde\kappa}
\newcommand{\ks}{{k}}
\newcommand{\bx}{\mb x}
\newcommand{\br}{\mb r}
\newcommand{\bu}{\mathbf u}
\newcommand{\bD}{\mathbf D}
\newcommand{\bA}{\mathbf A}
\newcommand{\bB}{\mathbf B}
\newcommand{\bC}{\mathbf C}
\newcommand{\bp}{\mathbf p}
\newcommand{\bq}{\mathbf q}
\newcommand{\by}{\mathbf y}
\newcommand{\pdg}{\bp\cdot\nabla}
\newcommand{\pdgx}{\bp\cdot\nabla_\bx}
\newcommand{\one}{1\hspace{-4.4pt}1}
\newcommand{\corr}{r_{\eta,\rho}}
\newcommand{\rinf}{r_{\eta,\infty}}
\newcommand{\rzero}{r_{0,\rho}}
\newcommand{\rzeroinf}{r_{0,\infty}}

\nwc{\nwt}{\newtheorem}

\nwt{remark}{Remark}
\nwt{definition}{Definition} 

\nwc{\ba}{{\mb a}}
\nwc{\bal}{\begin{align}}
\nwc{\be}{\begin{equation}}
\nwc{\ben}{\begin{equation*}}
\nwc{\bea}{\begin{eqnarray}}
\nwc{\beq}{\begin{eqnarray}}
\nwc{\bean}{\begin{eqnarray*}}
\nwc{\beqn}{\begin{eqnarray*}}
\nwc{\beqast}{\begin{eqnarray*}}

\nwc{\eal}{\end{align}}
\nwc{\ee}{\end{equation}}
\nwc{\een}{\end{equation*}}
\nwc{\eea}{\end{eqnarray}}
\nwc{\eeq}{\end{eqnarray}}
\nwc{\eean}{\end{eqnarray*}}
\nwc{\eeqn}{\end{eqnarray*}}
\nwc{\eeqast}{\end{eqnarray*}}

\nwc{\ep}{\varepsilon}
\nwc{\eps}{\varepsilon}
\nwc{\ept}{\epsilon}
\nwc{\vrho}{\varrho}
\nwc{\orho}{\bar\varrho}
\nwc{\ou}{\bar u}
\nwc{\vpsi}{\varpsi}
\nwc{\lamb}{\lambda}

\nwt{proposition}{Proposition}
\nwt{theorem}{Theorem}
\nwt{summary}{Summary}
\nwc{\nn}{\nonumber}
\nwc{\mf}{\mathbf}
\nwc{\mb}{\mathbf}
\nwc{\ml}{\mathcal}

\nwc{\IA}{\mathbb{A}} 
\nwc{\IB}{\mathbb{B}}
\nwc{\IC}{\mathbb{C}} 
\nwc{\ID}{\mathbb{D}} 
\nwc{\IM}{\mathbb{M}} 
\nwc{\IP}{\mathbb{P}} 
\nwc{\II}{\mathbb{I}} 
\nwc{\IE}{\mathbb{E}} 
\nwc{\IF}{\mathbb{F}} 
\nwc{\IG}{\mathbb{G}} 
\nwc{\IN}{\mathbb{N}} 
\nwc{\IQ}{\mathbb{Q}} 
\nwc{\IR}{\mathbb{R}} 
\nwc{\IT}{\mathbb{T}} 
\nwc{\IZ}{\mathbb{Z}} 

\nwc{\cE}{{\ml E}}
\nwc{\cP}{{\ml P}}
\nwc{\cQ}{{\ml Q}}
\nwc{\cL}{{\ml L}}
\nwc{\cX}{{\ml X}}
\nwc{\cW}{{\ml W}}
\nwc{\cZ}{{\ml Z}}
\nwc{\cR}{{\ml R}}
\nwc{\cV}{{\ml L}}
\nwc{\cT}{{\ml T}}
\nwc{\crV}{{\ml L}_{(\delta,\rho)}}
\nwc{\cC}{{\ml C}}
\nwc{\cA}{{\ml A}}
\nwc{\cK}{{\ml K}}
\nwc{\cB}{{\ml B}}
\nwc{\cD}{{\ml D}}
\nwc{\cF}{{\ml F}}
\nwc{\cS}{{\ml S}}
\nwc{\cM}{{\ml M}}
\nwc{\cG}{{\ml G}}
\nwc{\cH}{{\ml H}}
\nwc{\bk}{{\mb k}}
\nwc{\cbz}{\overline{\cB}_z}
\nwc{\supp}{{\hbox{supp}(\theta)}}

\nwc{\pft}{\cF^{-1}_2}

\title{Time Reversal of Broadband Signals in a Strongly Fluctuating MIMO Channel: Stability and Resolution}
\author{Albert Fannjiang}
 \email{
  cafannjiang@ucdavis.edu}
 
 \thanks{
 The research supported in part by  NSF grant DMS 0306659,  DARPA Grant N00014-02-1-0603.}
 \address{
Department of Mathematics,
University of California, Davis, CA 95616-8633}

\maketitle

\noindent
{\bf 
We analyze the time reversal of a multiple-input-multiple-output
(MIMO)
system in a space-frequency-selective multi-path fading 
channel described by the stochastic Schr\"odinger equation
with a random potential in the strong-fluctuation regime.
We prove that in a broadband limit the conditions for stable super-resolution are 
the {\em packing} condition  that the spacing among the $N$ transmitters
and $M$ receivers be more than 
the coherence length $\ell_c$  and the consecutive symbols in the data-streams
are separated by more than the inverse of the bandwidth $B^{-1}$ and the {\em multiplexing} condition that 
the number of the degrees of freedom per unit time at the transmitters ($\sim NB$) be much larger than
 the number of the degrees of freedom ($\sim MC$) per unit time in
the ensemble of intended messages. Here
$C$ is the number of symbols per unit time in the data-streams
intended for each receiver.
When the two conditions are met, 
all receivers receive simultaneously  streams of statistically stable, sharply focused
 signals intended for them, free of fading and interference. 
This indicates the rough  multiplexing gain of $NB$ in channel capacity, with the maximal gain per unit  {\em angular} cross section
given by $BL^d \ell_c^{-d}$ where $L$ is the distance from
the transmitters to the receivers.  We show that under
the ideal packing condition  time reversal can result in a high signal-to-interference ratio
 and low probability of intercept, and hence
 is an effective means for achieving the information capacity
 of  multi-path channels in the presence of multiple
 users (receivers).
}
\section*{Introduction}
Time reversal (TR) of waves is the process of recording
the signal from a remote source and then retransmitting
the signal in a time-reversed fashion to refocus on
the source (see \cite{Fink} and the references therein).
The performance of TR depends on, among other factors,  the reciprocity
(or time symmetry) of the propagation channel.
One of the most striking features of time reversal
operation in a strongly scattering medium is
super-resolution, the counterintuitive effect of
scattering-enhancement of time reversal resolution 
\cite{DTF2}, \cite{BPZ}, \cite{tire-phys}. It highlights
the great potential of time reversal in technological
applications such as communications where
the ability of steering and pinpointing signals 
is essential for realizing the information carrying
capacity of  a multi-path channel as well as
achieving  low probability
of intercept \cite{DTF}, \cite{KKP}. 

 In order to take full advantage of the super-resolution 
effect in a random medium, one has to first achieve statistical stability which can be measured by 
the signal-to-interference ratio (SIR) and
the signal-to-sidelobe ratio (SSR).
Statistical stability and resolution are two closely
related issues that should be analyzed side-by-side;
together, they are the measure of  performance of TR
which depends on, but
is not guaranteed by,  the reciprocity
(or time symmetry) of the propagation channel.
  It has been demonstrated experimentally 
that there are at least two routes to achieving statistical
stability \cite{DTF1}, \cite{DTF2}. One route is to use a time-reversal array (TRA) of sufficiently large aperture; the other is to use a broadband signal
(even with one-element TRA of essentially zero aperture).
There has been many advances in analytical understanding
of the former situation (see \cite{rad-arma} and
references therein). In many interesting applications
of time reversal, however,  the aperture
of TRA is typically small compared to the correlation length
of the medium, therefore the technological potential
of time reversal hinges  more heavily on the second route to
statistical stability. Compared to the case of large aperture the  analytical understanding of the case of broadband signals in time reversal has been so far
much less complete with the exception
of a randomly {\em layered} medium \cite{BPZ}.

In this paper we present the time reversal analysis
for the MIMO broadband channel whose $\ks$-component is described
by the stochastic Schr\"odinger equation \beq
&i\frac{\partial \Psi_z}{\partial z} +
\frac{\gamma}{2\ks} \Delta_\bx\Psi_z+\frac{\ks}{\gamma}
\chi_z\circ\Psi_z=0,\quad\bx\in \IR^d,\label{para}
\label{para1}
\label{para2}
\eeq
in the so called paraxial Markov approximation.
Here the refractive index fluctuation
$\chi_z(\cdot)$ is  a $\delta$-correlated-in-$z$ stationary random field with a power spectral density $\Phi(\bp)$
such that 
$
\IE\lt[\chi_z(\bx) \chi_{z'}(\bx')\rt]
=\delta(z-z')\int \Phi(\bp) e^{i\bp\cdot(\bx-\bx')}d\bp
$
with  $\IE$ standing for the ensemble average;
$\ks$ is the (dimensionless) relative wavenumber to
the center wavenumber $k_0$; 
the Fresnel number $\gamma=L_z/(k_0L_x^2)$ 
is a dimensionless number constituting of
the center wavenumber $k_0$ and
the reference scales $L_z$ and $L_x$ in
the longitudinal and transverse dimensions, respectively,
see FIG. 1.
The notation $\circ$ in eq. (\ref{para}) means
the Stratonovich product (v.s. It\^o product). 
For simplicity of presentation we will assume
isotropy, i.e. $\Phi(\bk)=\Phi(|\bk|),\forall \bk\in \IR^d$
and smoothness of $\Phi$. 

\begin{figure}
\begin{center}
\includegraphics[width=6.5cm, totalheight=4cm]{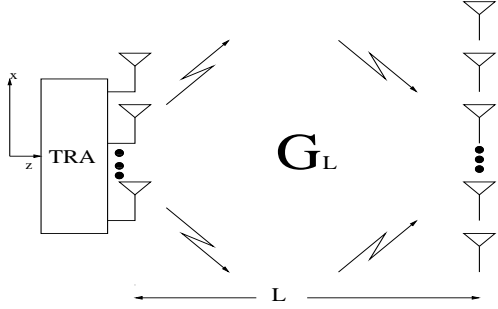}
\end{center}
\label{figback}
\caption{MIMO-TRA
 }
 \end{figure}

The stochastic  parabolic wave equation (\ref{para}) is a fundamental equation for wave
propagation in a randomly inhomogeneous {\em continuum} such as underwater
acoustic and electromagnetic waves in  atmospheric turbulence \cite{Ish}, \cite{TIZ}  as an approximation to the wave equation with random coefficients when 
backscattering and depolarization are weak.  It also models the cross-phase-modulation in nonlinear optical fibers
in the wavelength-division-multiplexing scheme \cite{Kaz}. 
It has a certain degree of universality and encapsulates the
{\em spatial } aspect of wave physics, a primary
focus of the present work, which is lacking in analysis
based on either
 randomly layered media
or random matrix theory \cite{Tel}, \cite{FG}, \cite{SM}.
Here  eq. (\ref{para}) is
treated as a model equation for  continuous random media
and is studied rigorously here to extract useful
insights that may shed light on other types of complex media.

Our goal is to show that for the channel described by (\ref{para}) the stability and super-resolution
can be achieved simultaneously when there is
sufficiently high number of degrees of freedom
at the transmitters.
In particular, we analyze the multiple-input-multiple-output (MIMO)
time-reversal communication satisfying 
 the {\em packing} condition 
 that, roughly speaking, the spacing of the $M$ receivers and $N$ elements of the time-reversal array (TRA) 
 is much more than the coherence length 
of the channel, and that the  consecutive symbols in the $T$-datum streams  are separated by more than $B^{-1}$, the inverse
of the non-dimensionalized frequency bandwidth  $B$($=\hbox{frequency bandwidth}$$\times L_x^2/L_z $).
Our main theorem says that 
in the {\em strong-fluctuation} regime and the
{\em broadband} limit (\ref{uwb}) the MIMO-TRA system achieves stable super-resolution in the sense that both the SIR
and SSR tend to infinity and that
 the  signal received by each receiver is  focused to
 within a circle of the coherence length $\ell_c$
when the additional {\em multiplexing } condition is also met, namely   $NB\gg MT\beta_c$ where $\beta_c$ is the coherence
bandwidth.

To further understand the meaning and implication of the result, 
we may assume without loss of generality that
the timing of the data-streams are within 
one interval of the delay spread since the signals separated by much more than one delay spread are roughly uncorrelated.
Because the delay spread $\delta_*\sim\beta_c^{-1}$ (cf. the section entitled {\bf From SISO to MIMO}), $T\beta_c$ is roughly the
number of symbols per unit time in each data-stream and hence $MT\beta_c$ is the number of the degrees of freedom 
per unit time in 
the ensemble of all data-streams  while $NB$ is
the total number of the degrees of freedom per unit time at the TRA. The multiplexing condition says
that the number of degrees of freedom of the intended
messages must be smaller than the number of degrees
of freedom available in the channel. The main technical ingredient of our approach is the {\em exact},
{\em universal}
low Fresnel number 
asymptotic obtained for the two-frequency
mutual coherence function. The calculation is tight
indicating that the  multiplexing condition  is sharp.

The main assumption is the 4-th order sub-Gaussianity property (\ref{gauss2}). 
The Gaussian-like behavior for 4-th order correlations is widely believed to occur in the {\em strong-fluctuation} regime, defined
 by $\alpha_*^2=D_2L\gg 1, \sigma_*^2=D_2L^3\gg 1$.
We will point out some independent evidences for this in our
calculation.
Here 
$L$ is the (longitudinal) distance between the TRA and the receivers and $D_2=d^{-1}\int |\bp|^2\Phi(\bp)d\bp$ is the {\em angular}
diffusion coefficient (hence $\alpha_*=\sqrt{D_2L}$ is the angular spread).
In the strong-fluctuation regime \cite{tire-phys}, $\ga^{-1}\alpha_*$ is
the spread in the so called {\em spatial frequency}, $\sigma_*=\sqrt{D_2L^3}$
 the spatial spread and  their product $\ga^{-1}D_2L^2$  
the {\em spatial-spread-bandwidth} product (SSB)
which, as we will show, is exactly $\ga^{-1}\beta_c^{-1}$.
\commentout{
Raised to the power $d$, SSB is roughly the phase-space
volume occupied by the wave energy of the  Green function at $\ks=1$,
 measured in the natural unit $\ga^d$ dictated by the uncertainty principle on the phase space.
 }
By the duality principle for the strong-fluctuation regime,  proved in \cite{tire-phys}, the {\em effective aperture} is $2\pi$ times the  spatial spread $\sigma_*$
(independent of the numerical aperture of TRA and hence super-resolution),  and  its 
 dual quantity $\gamma L/(\ks\sigma_*)\approx\gamma/\alpha_*$ (the inverse of spatial-frequency spread) is
 the coherence length $\ell_c$ of the forward propagation
 (as well as the time reversal resolution). 
Hence the ratio $ \sigma_*/\ell_c$ equals the spatial-spread-bandwidth product  and is roughly the number of
uncorrelated sub-channels (paths)  per transverse dimension in the cross section of diameter $\sigma_*$ at the receiver plane,
which will place upper bound on the capacity gain per unit angle 
of the channel (see more on
this in the Conclusion).

 In what follows, we first formulate the problem and develop the essential tool for
 analyzing TR, the 
 one- and two-frequency mutual coherence functions, and
 then  carry out  the stability and resolution analysis for the single-input-single-output (SISO),  multiple-output-single-output (MISO),
 single-input-multiple-output (SIMO) and the multiple-input-multiple-output (MIMO) cases. Both MISO- and SIMO-TRA systems have been demonstrated
 to be feasible for ocean acoustic communication \cite{RJD}, \cite{EK}, \cite{KK} and the MIMO-TRA system with $N>M$ has been shown to work
 well for ultrasound  \cite{DTF}. We will discuss the implications
 of our results
 on the channel capacity in the Conclusion.
 We have by and large neglected the effect of noise in our
 analysis, assuming that the TRA operates in a high
 signal-to-noise ratio (SNR) situation as is the case
 for  the experiments reported in \cite{DTF}, \cite{EK}.
 The robustness of TR in the presence of noises has been well
 documented, see e.g. \cite{Dow}.
 
\section*{MIMO-Time reversal}

We extend the time-reversal
communication scheme \cite{DTF} to 
the setting with multiple users. Let the $M$
receivers  located at $(L,\br_j), j=1,...,M$ first send a pilot signal $\int e^{i\frac{\ks t}{\ga}}g(\ks) d\ks \delta(\br_j-\ba_i)$ to
the $N$-element TRA located at $(0,\ba_i), i=1,...,N$ which then use the time-reversed
version of the received signals $\int e^{i\frac{\ks t}{\ga}}g(\ks) G_L(\br_j,\ba_i;\ks)d\ks$ to modulate streams of symbols and send them
back to the
receivers. Here  $G_L$ is the Green function of eq. (\ref{para})
and  $g^2(\ks)$ is the power
density at $\ks$.
As shown in  \cite{BPZ},  \cite{DLF},    when the TRA has an infinite time-window
(see the Conclusion for the case of finite time-window),
 the signal arriving at the receiver plane  with delay $L+t$ is given by 
  \beq
 \nn
{S(\br,t)}
&= &\sum_{l=1}^T\sum_{i=1}^N\sum_{j=1}^M m_j(\tau_l)\int  e^{-i\frac{\ks}{\ga}(t-\tau_l)}g(\ks)\\
&&\times
G_L(\br,\ba_i;\ks)G^{ *}_L(\br_j, \ba_i;\ks)d\ks
\label{mr}
\eeq
where $m_j(\tau_l), l=1,...,T\leq\infty$ are a stream of $T$ symbols intended for the $j$-th
receiver transmitted at times $\tau_1<\tau_2<...<\tau_T$.
We assume for simplicity that $|m_j(\tau_l)|=1$, $\forall j, l$. We have chosen the time scale such that the speed
of propagation is unity (thus wavenumber=frequency). 

 We assume  that $g$ is a smooth and rapidly decaying
function with effective support of size $B\ga$. For simplicity
we take $g^2(\ks)=\exp{(-\frac{|\ks-1|^2}{2B^2\ga^2})}$.  The broadband limit 
may be formulated as the double limit 
\beq
\label{uwb}
\gamma\to 0,\quad  B\to \infty
\quad
\lim B\gamma=0
\eeq
so that in the limit $g^2(\ks)$ becomes narrowly
focused around $\ks=1$.
The idea underlying the definition is to view
the broadband limit as a sequence of narrow-bands
with indefinitely growing center frequency and bandwidth.
This is particularly well suited to the framework
of parabolic approximation described by (\ref{para1}).
The apparent narrow-banding of (\ref{uwb}) is
deceptive: the {\em delay-spread-band-width} product (DSB)
turns out to be $B\beta_c^{-1}$ and is  doubly divergent 
as $B\to \infty$ (the broadband limit) and
$\beta_c\to 0$ (the strong fluctuation regime).  Note that
since  $\ks$ is the {\em relative} wavenumber,  the product $B\gamma$ should
always be  uniformly bounded between zero and unity, independent of $\gamma>0$. In the case $d=1$ this 
has the intuitive implication that the number $B\beta^{-1}_c$ of degrees
of freedom at each TRA-element  is less than or equal to
the number  $\ga^{-1}\beta_c^{-1}$ (SSB) of uncorrelated propagation paths in the medium.

{\bf Packing condition.}
\commentout{
 We assume   that the $N$ TRA-elements
be separated by  much more than
$\ell_c\sqrt{\log{(1+N)}}$,
 the  $M$
receivers  be separated by much more than 
$\ell_c\sqrt{\log{(1+M)}}$ from one another and  the  consecutive symbols $\tau_l, \tau_{l+1}$ be
separated by much more than $B^{-1}\sqrt{\log{(B\beta_c^{-1})}}$.  
} We assume that the spacing within the $N$ TRA-elements
and the $M$ receivers be much larger than the coherence
length $\ell_c$ and that
the separation of the successive symbols  be
much larger than $(2B)^{-1}$.
Though
there is no technical limitation on
$M,N,T$,  it suffices to consider the case
where all the $N$ TRA-elements and all the
$M$ receivers are located within one circle of  diameter $\sigma_*$
(implying $M,N\ll \ga^{-d}\beta_c^{-d}$), and all the $T$-datum streams are within
one interval of the delay spread $\sim \beta_c^{-1}$
(implying $T\ll B\beta_c^{-1}$) since the signals
separated by much more than one spatial spread $\sigma_*$
or one delay spread $\delta_*$ are essentially uncorrelated.

For simplicity, we have assumed that all the receivers
lie on the plane parallel to the TRA. When this is not
the case, then the above spacing of antennas refers
to the {\em transverse} separation parallel
to the TRA.

\commentout{
The square-root-log factors in the spacing of antennas and
the timing  of symbols are introduced in order to
achieve a clean-cut estimate for the variance of the signal
by using the 
Gaussian tails in $g$ and the mutual coherence functions
arising in the subsequent analysis.
}

{\bf SIR/SSR.} Anticipating a singular limit we employ the coupling with smooth, compactly supported  test functions.  Denote
the mean by $E(\br,t) = \ga^{-d}\int \theta^*((\bx-\br)/\ell_c)\IE S(\bx,t)d\bx$ where the coupling with the test function
$\theta$ can be viewed as the averaging induced by measurement.
Denote the variance by $V(\br,t)
=\ga^{-2d}\IE\big[\int \theta^*((\bx-\br)/\ell_c) S(\bx, t)d\bx  \big]^2-E^2(\br,t)$.
We have made  the test function $\theta$ act  on the  scale
of the coherence length $\ell_c$, the smallest spatial scale of interest
(the speckle size) in the present context. Different choices of scale
 would not affect the
conclusion of our analysis.

 The primary object of our analysis is
\beq
\label{rho}
\rho(\br,t)=
\frac{E^2(\br_j,\tau_l)}{V(\br,t)},\quad j=1,...,M, l=1,...,T
\eeq
which is  the SIR  if $\br=\br_j,t=\tau_l$ and the SSR if $|\br-\br_j|\gg \ell_c, \forall j$ (spatial sidelobes) or $|t-\tau_l|\gg B^{-1}, \forall l$ (temporal sidelobes)
(as $V(\br,\tau)\approx E^2(\br,\tau)$ as we will see below).
In the special case  of $\br=\br_j$ {\em and} $|t-\tau_l|\gg B^{-1},$$\forall l$, $\rho^{-1}$ is a measure of intersymbol
interference. 
To show stability and resolution, we shall find
the precise conditions under which 
 $\rho\to\infty$ and   
 $\IE S (\br,t)$ is asymptotically 
$
\sum_{l=1}^T \sum_{j=1}^Mm_j(\tau_l)S_{jl}(\br,t)
$
where $S_{jl}(\br,t)=0$ in the spatial or temporal sidelobes and 
\beq
\label{5.5}
\lefteqn{S_{jl}(\br,t)}\\
&\approx&\sum_{i=1}^N\int e^{-i\frac{\ks (t-\tau_l)}{\ga}}g(\ks)\IE\big[G_L(\br,\ba_i;\ks)
 G^{ *}_L(\br_j, \ba_i;\ks)\big] d\ks\nn
\eeq
is a sum of $\delta$-like functions around   $\br_j$ and $\tau_l=0, \forall l$.  In other words, we employ
  the TRA as  a multiplexer to transmit the
$M$ scrambled data-streams to the receivers
and we hope to
 turn the medium into a demultiplexer
  by employing the broadband time reversal technique.

\section*{Mutual coherence functions} 
A quantity repeatedly appearing in the subsequent analysis is
 the  mutual coherence  function $\Gamma_z$ 
 between the Green functions
at two different wavenumbers
  $\ks_1=\ks-\ga\beta/2,\ks_2= \ks+\ga\beta/2$ \beqn
{\Gamma_z(\frac{\bx+\br}{2},\frac{\bx-\br}{\ga};\ks,\beta)}
&=&\IE\big[G_{z}(\bx,\ba;\ks-\ga\beta/2)\\
&&\times G^*_{z}(\br,\ba';\ks+\ga\beta/2)\big].
\eeqn
We shall omit writing $\ks,\beta, \ba, \ba'$ when  no confusion arises.
Here we have chosen $\bx,\br$ to be the pair
of variables of concern and left out $\ba, \ba'$ as
parameters. By the reciprocity of the Green function,
we can choose one variable from $\{\bx,\ba\}$ and
the other from $\{\br,\ba'\}$ as the variables of $\Gamma_z$
and leave the others as parameters.

{\bf One-frequency version.} When $\beta=0$,  $\Gamma_z$ satisfies  
\beq
\label{ga1f}
\partz \Gamma_z+\frac{i}{\ks}\nabla_\bx\cdot\nabla_\by\Gamma_z
+\frac{\ks^2}{\gamma^2}D(\ga\by)\Gamma_z=0
\eeq
where  the structure function of the medium fluctuation $D(\bx)$ is given by
$
D(\bx)=\int \Phi(\bk)\lt[1-e^{i\bk\cdot\bx}\rt]d\bk\geq 0,\quad
\forall \bx\in \IR^d.
$
Eq. (\ref{ga1f}) is exactly solvable by the Fourier transform in $\bx$. For $\alpha_*\ll \ga^{-1}$ and $\sigma_*\ll \ga^{-1}$ we can use the approximation
\[
\ks^2\ga^{-2}\int^L_0D\lt(\ga\by-\bp z\gamma/\ks\rt)dz\approx
\int^L_0D_2|\ks\by-\bp z|^2dz
\]
to obtain 
\beq
\label{one}
\label{5.1}
\lefteqn{\Gamma_L(\bx,\by;\ks,0)}\\
&\approx& \int e^{i\bp\cdot\bx}\hat\Gamma_0(\bp,\by-\frac{L\bp}{\ks};\ks,0)
e^{-\int^L_0 D_2\lt|\ks\by+\bp z\rt|^2 dz} d\bp\nn\\
&=& \int e^{i\bp\cdot\bx}\hat\Gamma_0(\bp,\by-\frac{L \bp}{\ks};\ks,0)
e^{-\int^1_0 \lt|\tilde\by+\tilde\bp z\rt|^2 dz} d\bp\nn
\eeq
where $\tilde \by={\by}\alpha_*\ks$
and $\tilde\bp=\bp\sigma_*$
It is clear from (\ref{5.1}) that $\Gamma_L$ has
a Gaussian-tail in $\by$ (the difference coordinates)
and, by rescaling,
an effective support  $\sim \alpha_*^{-1}$ and hence
$\ell_c=\ga/\alpha_*$ (recall that $\by$ is the coordinate
on the scale $\ga^{-1}$).

{\bf Two-frequency version.} 
The two-frequency mutual coherence function is  not exactly solvable except for some special cases. Fortunately the asymptotic for $\gamma\ll 1$ has
a universal form and  can be calculated exactly.
Without loss
of generality we assume $\beta>0$ in what follows.

Using the so called two-frequency Wigner distributions we  have proved in   \cite{2f-whn} that  in the limit $\gamma\to 0$,
$\Gamma_z$ satisfies
the equation
\beq
\label{mean-eq}
{\frac{\partial\Gamma_z}{\partial z}
-\frac{i}{k}\nabla_\by\cdot\nabla_\bx \Gamma_z}
&=&-D_2\lt|\ks\by+\frac{\beta}{2}\bx\rt|^2
  \Gamma_z
-\frac{\beta^2}{2}D_0\Gamma_z
\eeq
where $D_0=\int\Phi(\bk)d\bk$.
The key to understanding eq. (\ref{mean-eq})
is the rescaling: 
\beq
\label{new}
\tilde\bx=\frac{\bx}{\sigma_*},\,\,
\tilde\by=\by\ks\alpha_*,\,\,\tilde z=z/L,\,\,
\tilde\beta={\beta}/{\beta_c}\eeq
with $\beta_c={D^{-1}_2 L^{-2}}
$
 which transforms eq. (\ref{mean-eq}) into the form
\beq
\label{mean-eq2}
{\frac{\partial\Gamma_z}{\partial \tilde z}
-{i}\nabla_{\tilde\by}\cdot\nabla_{\tilde\bx} \Gamma_z}
&=&-\big|\tilde\by+\frac{\tilde\beta}{2}\tilde\bx\big|^2
  \Gamma_z
-\frac{\tilde\beta^2D_0\Gamma_z}{2\sigma_*^2}
\eeq

By another change of variables
$
\bz_1=\tilde\by+{\beta}\tilde\bx/2,\,\,
\bz_2=\tilde\by-{\beta\tilde\bx}/2$
eq. (\ref{mean-eq2}) is then transformed into 
that of the {\em quantum harmonic oscillator} 
and solved exactly.
The solution is given by
\beq
\label{gamma}
\lefteqn{\Gamma_L(\bx,\by;\ks,\beta)}\\
\nn&=&
\frac{(2\pi)^d(1+i)^{d/2}\tilde\beta^{d/4}}{\sin^{d/2}{\big(\tilde\beta^{1/2}(1+i)\big)}} e^{-\frac{\tilde\beta^2D_0}{2\sigma_*^2}}\int d\bx' d\by' 
e^{i\frac{
|\tilde\by-\by'|^2}{2\tilde\beta}}
\\
\nn&&\times e^{\frac{1-i}{2\sqrt{\tilde\beta}}
\cot{(\sqrt{\tilde\beta}(1+i))}\big|
\tilde\beta\tilde\bx+\tilde\by
-\frac{\tilde\beta\bx'+\by'}{
\cos{(\sqrt{\tilde\beta}(1+i))}}\big|^2}
\\
\nn&&\times e^{-\frac{1-i}{2\sqrt{\tilde\beta}}
\lt|\tilde\beta\bx'+\by'\rt|^2
\tan{(\sqrt{\tilde\beta
}(1+i))}}\Gamma_0(\sigma_*\bx',\frac{\by'}{\ks\alpha_*}).\nn
\eeq
Several  remarks are in order:
(i) The Green function for  $\Gamma_L$ is of the Gaussian form
in $\bx,\by$,
consistent with the (sub-)Gaussianity assumption;
(ii) In the vanishing fluctuation limit $D_0,  D_2\to 0$ the 
free-space two-frequency mutual coherence function  is recovered;
(iii)  The apparent singular nature of the limit
$\beta\to 0$ in (\ref{gamma}) is deceptive. Indeed,
the small $\beta$ limit is regular and yields the result
 obtained from eq. (\ref{mean-eq}) with $\beta=0$;
 (iv) In the strong-fluctuation regime, $D_0$ is typically
 much smaller than $D_2^2L^3\gg 1$ so the factor
 $\exp{\big(-\tilde\beta^2 D_0/(2\sigma_*^2)\big)}$
 is negligible in the strong-fluctuation regime. On the
 other hand, the rapidly decaying factor $\sin^{-d/2}{\big(\tilde\beta^{1/2}(1+i)\big)}$ is crucial
 for the stability argument below;
(v)  $\Gamma_L(\bx,\by;\ks,\beta)$ 
is slowly varying in $\bx$ on
the scale $\sigma_*$ for $\beta\sim\beta_c$
and
more rapidly varying in $\bx$ for $\beta\gg \beta_c$.

{\bf Fourth order sub-Gaussianity.}
The strong-fluctuation regime $\alpha_*\gg 1, \sigma_*\gg 1$ can result from  either
long distance propagation and/or large medium fluctuation.   It is widely accepted that, in this regime, the statistics of the wave fields (for at least lower moments) become Gaussian-like resulting in, for $d=2$, an exponential PDF for the intensity 
\cite{Goo}, \cite{TZ}, \cite{TIZ}, \cite{FPS}, \cite{Sh}, \cite{SS}.
The Gaussian
statistics follows heuristically from Central-Limit-Theorem as
the number of uncorrelated sub-channels (paths) per transverse dimension in the cross section of diameter $\sigma_*$
increases linearly with the spatial-spread-bandwidth product,
as explained in the Introduction.
This 
 is consistent with the experimental finding of
the saturation of intensity fluctuation  with the scintillation index 
approaching unity \cite{Ish}.

In what follows
we shall make the 4-th order {\em sub-Gaussianity} hypothesis, namely 
that the fourth moments of the Green function  at different frequencies $\{G_L(\ks)\}$ can be estimated
by those of the
 Gaussian process of the same covariance. More specifically,
 we assume 
 that 
 \beq
 \label{gauss2}
&&\big|\IE\lt[G_L(\ks_1)\otimes G^*_L(\ks_1)\otimes G_L(\ks_2)
\otimes G^{ *}_L(\ks_2) \rt]\\
\nn&&-
\IE\lt[G_L(\ks_1)\otimes G^*_L(\ks_1)\rt]\otimes
\IE\lt[G_L(\ks_2)\otimes
G^{ *}_L(\ks_2) \rt]\big|
\\
\nn&\leq& K
\big|\IE\lt[G_L(\ks_1)\otimes 
G_L(\ks_2)\rt]\otimes \IE\lt[G_L^*(\ks_1)\otimes
G_L^*(\ks_2)\rt]\big|\\
\nn&& +K\big|
\IE\lt[G_L(\ks_1)\otimes
G^*_L(\ks_2)\rt]\otimes\IE\lt[G_L^*(\ks_1)\otimes 
G_L(\ks_2)\rt]\big|
\eeq
for some constant $K$ independent of $\ga\to 0, |\ks_1-1|=O(B\ga), |\ks_2-1|=O( B\ga)$ and all the variables.  For a jointly Gaussian process,
the constant $K=1$. Note that, in view of
the scaling in the two-frequency mutual coherence the first term on the RHS of
(\ref{gauss2}) is much smaller than the second term
due to difference in wavenumber for $G_L(\ks)=G_L^*(-\ks)$.

The sub-Gaussianity assumption will be used
to  estimate the 4-th order correlations of
Green functions appearing in the calculation for $V$ by the two-frequency mutual coherence function
in the strong-fluctuation regime.

\section*{From SISO to MIMO}
\label{mm}
Our first application of the mutual coherence functions
is the estimate for the delay spread.
Consider the band-limited impulse response $u(\bx,t)=$$\int g(\ks)$$
e^{\frac{i\ks (L-t)}{\ga}}$$ G_L(\bx,0;\ks)d\ks$. It follows
easily using the preceding results that the mean delay is $L$ 
 and the asymptotic for the delay spread $\delta_*$, when $B\gg\beta_c$,
 is given by 
\beqn
\delta_*&=&\sqrt{\int (t-L)^2\IE | u(\bx,t)|^2dt/\int \IE|u(\bx,t)|^2 dt}\\
&\approx& \sqrt{-\frac{d^2}{d\beta^2}\Big|_{\beta=0}\Gamma_L(\bx,0;1,\beta)/\Gamma_L(\bx,0;1,0)}
\sim \beta_c^{-1}
\eeqn
which is slowly varying in $\bx$ on the scale $\sigma_*$.
As commented before it suffices to consider
the case with  a finite $T$ such that $|\tau_1-\tau_T|\sim \beta_c^{-1}$, implying the number of symbols in each data-stream $T \ll B\beta_c^{-1}$, the DSB. In what follows, due to $\beta_c\ll 1$ the temporal component of the signals  is essentially decoupled from
the spatial component and 
determined  by the power distribution $g^2$.

{\bf SISO. }
This case corresponds to $N=1, M=1$. Let $\ba_1=0$.
In the calculation of $E(\bx,t)$, the expression \[
\lan\theta,\Gamma_L\ran(\br)
\equiv\int \theta^*(\frac{\br_1-\br}{\ell_c}+\frac{\by\ga}{\ell_c})\Gamma_L(\br_1+\frac{\by\ga}{2},\by;\ks,0)d\by
\]
 arises and involves only the one-frequency
mutual coherence.  Using (\ref{one}) with $\Gamma_0(\bx,\by)=\delta(\bx+\frac{\ga\by}{2})
\delta(\bx-\frac{\ga\by}{2})$ and making the necessary rescaling of variables we obtain
the following asymptotic
\beq
\label{5.10}
&&{\lan\theta,\Gamma_L\ran(\br)}
\approx  C_0(\br,\br_1)\beta_c^{d}\\
&&C_0=
\int d\bp \theta^*({\bp}+\frac{\br_1-\br}{\ell_c})e^{-\frac{i\bp\cdot\br_1}{\sigma_*}}e^{-|\bp|^2/3}.\label{5.102}
\eeq
To derive (\ref{5.10}) we have used
the defining conditions  of the strong-fluctuation regime. 
Note that the transfer function in
(\ref{5.102})
is Gaussian in $\bp$ and that $C_0(\br,\br_1)$ has a Gaussian-tail in $|\br-\br_1|/ \ell_c$ and 
$C_0(\br_1,\br_1)$ is bounded away from zero and slowly varying in $\br_1$ on the scale  $\sigma_*$.  That is, after proper
normalization 
$C_0(\br,\br_1)$ behaves like a $\delta$-function
centered at  $\br_1$. By (\ref{5.10})-(\ref{5.102}) we obtain  the mean field asymptotic    $E(\br,t)\approx 0$ for
$|\br-\br_1|\gg \ell_c$ (spatial sidelobes) or $|t-\tau_l|\gg B^{-1}, \forall l$ (temporal sidelobes) and 
$
E(\br_1,\tau_l)\approx \sqrt{4\pi} C_0(\br_1,\br_1)
 \beta_c^dB\ga.
 $

The calculation for $V$ involves
the four-point correlation of the Green functions at different frequencies.
Under the sub-Gaussianity condition (\ref{gauss2}) the calculation reduces to that of two-frequency mutual coherence functions.

Using (\ref{gamma})   with $\Gamma_0(\bx,\by)=\delta(\bx+\frac{\ga\by}{2})
\delta(\bx-\frac{\ga\by}{2})$ we obtain  the asymptotic
for the dominant term  in the calculation for $V(\bx,\tau)$
prior to the $\ks$-integration
\beq
&&\hspace{-.5cm}\Gamma_L(\br_1,0;\ks,\beta)\int\Gamma_L(\frac{\bx_1+\bx_2}{2},\frac{\bx_1-\bx_2}{\ga};\ks,\beta) \nn\\
&&\times \theta^*(\frac{\bx_1-\br}{\ell_c})\theta(\frac{\bx_2-\br}{\ell_c})d\frac{\bx_1}{\ga} d\frac{\bx_2}{\ga}\approx C_{\tilde \beta}
 \beta_c^{2d}
\label{5.22}
\eeq
with the constant $C_{\tilde\beta}$  given by
\beqn
C_{\tilde\beta}&=&
{(2\pi)^{2d}(1+i)^d\tilde\beta^{d/2}}{\sin^{-d}{(\sqrt{\tilde\beta}(1+i))}}e^{-\frac{\tilde\beta^2D_0}{\sigma_*^2}}\\
&&e^{\frac{(1-i)}{2\sqrt{\tilde\beta}}\cot(\sqrt{\tilde\beta}(1+i))
\frac{\tilde\beta^2|\br_1|^2}{\sigma_*^2}}\int \theta^*(\tilde\by+\frac{\tilde\by'}{2})\theta(\tilde\by-\frac{\tilde\by'}{2})
\nn\\
&&\times
e^{\frac{i}{2\tilde\beta}|
\tilde\by'|^2}e^{\frac{1-i}{2\sqrt{\tilde\beta}}
\cot{\lt(\sqrt{\tilde\beta}(1+i)\rt)}\big|
\frac{\tilde\beta \br}{\sigma_*}+\tilde\by'\big|^2}d\tilde\by d\tilde\by'
.\nn
\eeqn

Due to the rapidly decaying factor $\sin^{-d}{\big(\sqrt{\tilde\beta}(1+i)\big)}$  the $\tilde\beta$-integration of $C_{\tilde\beta}$ is  convergent as $B\to \infty$.
Because $\beta_c\ll 1$, in the $(\ks_1,\ks_2)$-integration the power distribution $g(\ks_1)g(\ks_2)$
and $C_{\tilde\beta}$ are decoupled after the
change of variables: $(\ks_1,\ks_2)=(\ks-\beta\ga/2,\ks+\beta\ga/2)$, so  
we conclude that
$
V(\bx,t)$$
\leq$$ 2\sqrt{2\pi} K\ga^2$$\beta_c^{2d+1} BT\int C_{\tilde\beta}d\tilde\beta.
$
Note that the variance increases linearly with the number $T$ of symbols in each  data-stream.

The asymptotic SIR/SSR for the SISO-TRA is given by
$\rho=O\lt(B\beta^{-1}_cT^{-1}\rt)$.
Note that the SIR/SSR is slowly varying in the test point $\br$
and the receiver location $\br_1$  on the scale of $\sigma_*$. 

{\bf SIMO.} Let us turn to the SIMO case with $N=1$ element TRA located at 
$\ba_1=0$. 

The mean field calculation is analogous to the SISO case. Namely,
$E(\br_j,\tau_l)\approx\sqrt{4\pi}C_0(\br_j,\br_j)
 \beta_c^dB\ga$ and zero in the temporal or spatial
 sidelobe.

In view of the  the remark following (\ref{gauss2}) the variance of $S$ is dominated by the contribution from  the diagonal terms in the summation over receivers given by
\beqn
&&\sum_{l,l'=1}^T\int e^{-\frac{\ks(\tau_l-\tau_{l'})}{\ga}}g^2(\ks)d\ks \sum_{j=1}^M\int \theta^*(\frac{\bx_1-\br}{\ell_c})\theta(\frac{\bx_2-\br}{\ell_c})\\
&&\times\int d\beta \Gamma_L(\br_j,0;\ks,\beta)\Gamma_L(\frac{\bx_1+\bx_2}{2},\frac{\bx_1-\bx_2}{\ga};\ks,\beta)d\frac{\bx_1}{\ga}d\frac{\bx_2}{\ga} \\
&&\approx \sqrt{2\pi}B\ga^2 TM \int C_{\tilde \beta}d\tilde\beta
 \beta_c^{2d},
\eeqn
because 
$|\br_i-\br_j|\gg \ell_c$
regardless 
whether the test point is near or away from any receiver.
Therefore we have the estimate: $
\rho
=O\lt(B\beta^{-1}_cM^{-1}T^{-1}\rt).
$

{\bf MISO. }
The  case corresponds to  $M=1$. Each term
in the summation over the $N$ TRA-elements
has the same asymptotic as 
that of the SISO case. Hence
$E(\br_j,\tau_l)\approx \sqrt{4\pi}N C_0(\br_j,\br_j)\beta^{d}_c B\ga$
and zero in the spatial or temporal sidelobes.

For the variance calculation, let us first note that the correlations of two Green functions starting
with two TRA-elements located at $\ba_i, \ba_j$
satisfy eq. (\ref{mean-eq}) in the variables $(\ba_i,\ba_j)$, by the reciprocity of
the time-invariant channel, and hence vanish
as $|\ba_i-\ba_j|\gg\ell_c$.
  The variance of the signal at $\br$ (whether at $\br_1$ or away from it)   before performing  the $\ks$-integration is then dominated
  by the following diagonal terms in the summation over receivers
  \beqn
  &&\sum_{j=1}^N\IE\lt[G_L^*(\br_1,\ba_j;\ks_1)G_L(\br_1,\ba_j;\ks_2)\rt] \int \theta^*(\frac{\bx_1-\br}{\ell_c})\\
  &&\times\theta(\frac{\bx_2-\br}{\ell_c})\IE\lt[G_L(\bx_1,\ba_j;\ks_1)
  G_L^*(\bx_2,\ba_j;\ks_2)\rt]d\frac{\bx_1}{\ga} d\frac{\bx_2}{\ga}\\
  & &\approx NC_{\tilde\beta}\beta_c^{2d}. \eeqn
  The $\ks$-integration induces the additional factor of $\sqrt{2\pi}B\ga^2 T$. Hence
$
V(\br,t)$$
\leq$$ 2\sqrt{2\pi} K\ga^2$$\beta_c^{2d+1} BT\int C_{\tilde\beta}d\tilde\beta
$  since $|\br-\br_j|\ll \sigma_*, \forall j$.
We conclude that 
$\rho
=O\lt(NB\beta^{-1}_c T^{-1}\rt)$.

{\bf MIMO. }
The analysis for the MIMO case combines all the previous cases. 
The mean signal has the same asymptotic as that of the MISO case, i.e., linearly proportional to $BN$.
  The
variance of the signal prior to performing the $\ks$-integration is dominated
by
  \beqn
 &&\sum_{i,j=1}^{M,N}\IE\lt[G_L^*(\br_i,\ba_j;\ks_1)G_L(\br_i,\ba_j;\ks_2)\rt] \int \theta^*(\frac{\bx_1-\br}{\ell_c})\\
  &&\times\theta(\frac{\bx_2-\br}{\ell_c})\IE\lt[G_L(\bx_1,\ba_j;\ks_1)
  G_L^*(\bx_2,\ba_j;\ks_2)\rt]d\frac{\bx_1}{\ga} d\frac{\bx_2}{\ga}\\
  &&\approx NMC_{\tilde\beta}\beta_c^{2d} \eeqn
  and therefore
$V
\leq T MN2K\ga^2 B\beta^{2d+1}_c\int C_{\tilde\beta}d\tilde\beta.
$

  We collect the above analysis in the following statement.

{\bf Summary}. {\em Let the $N$-element TRA,
$M$ receivers and the number of symbols $T$  satisfy the packing condition.
Assume
the 4-th order sub-Gaussianity condition (\ref{gauss2}) in
the strong-fluctuation regime and let $
1\ll \alpha_*\ll \ga^{-1},1\ll \sigma_*\ll \ga^{-1}$.

Then in the broadband limit (\ref{uwb}) 
the asymptotic  SIR/SSR  $\sim NM^{-1}T^{-1}B\beta^{-1}_c$ is valid 
uniformly for all $\br_j, j=1,...,M$, with the constant of proportionality
$2^{-1} (2\pi)^{-1/2}K^{-1}(\int C_{\tilde\beta}d\tilde\beta)^{-1}|C_0|^{2}$ where $C_0$
as given by (\ref{5.102}) is not zero for $\theta\not \equiv 0$.

The asymptotic signal at the receiver plane 
within the distance $\ll$$\sigma_*$ from the receivers is $\sum_{l=1}^T\sum_{j}^M m_j(\tau_l) S_{jl}(\bx,t)$ where $S_{jl}(\bx,t)$ given by (\ref{5.5}).
}

\section*{Conclusion and discussion}

\label{last}
The strong-fluctuation regime constitutes 
the so called space-frequency-selective multi-path fading 
channels in wireless communications \cite{Paul}.
In such a channel, TR has the  super-resolution given by $\ell_c=\ga/\sqrt{D_2L}$. We have established
firmly the packing and multiplexing  conditions for stable super-resolution
for the MIMO-TRA communication system
under the 4-th order sub-Gaussianity assumption.
The experimental evidence for our result in the case of $M=1$ has
been demonstrated in \cite{DTF2}.

We have argued that statistical stability is crucial for multi-receiver TR communications, especially when
the multiple receivers do not have channel state information, as the multiuser interference
is essentially indistinguishable from the intended signal,
the only difference being their statistical properties.
The latter is in the mean field while the former
is primarily  in the fluctuating field. Our result implies
that the time-reversal communication can be realized
{\em stably} 
in principle with up to
 $M\sim NB\beta_c^{-1}T^{-1}$ receivers simultaneously
 at the rate $T\beta_c$ with {\em low probability of intercept} 
due to super-resolution.  Concerning the channel capacity,
 our result is analogous to the finding  
 in
\cite{FG}, \cite{Tel}, \cite{Mou}, \cite{SM} 
based on the random matrix modeling and theory
that the {\em ergodic} capacity with complete channel state information at
the receiver with $M$ receive antennas (but not at the 
$N$ transmit antennas) scales like $ \min{(M,N)}\log_2{\hbox{SNR}}$  (per unit frequency) at high SNR. 
 After taking into account
 the frequency multiplexing gain \cite{CT}, \cite{Tel}, the multi-frequency
 channel capacity then scales like
  $ B\min{(M,N)}\log_2{\hbox{SNR}}$. Note, however,
  this result does not include the interference due to noncooperating
  multiuser receivers as we do here. Also,
  these works consider only  narrow-band signals
  for which statistical stability is rarely valid
  in practice and consequently the ergodic capacity
  is an average, not almost sure,  quantity. 
    
In the present set-up with the $B$-band-limited channel state information at the transmitters but not the receivers, the multiplexing gain is, up to a logarithmic factor,  roughly $MT\beta_c\sim BN$,
the number of degrees of freedom per unit time
at TRA (see \cite{tr-comm} for more analysis on TR capacity
in multi-path Rayleigh fading channels). 
The packing condition also points to the maximal capacity  per
unit {\em angular } cross section $ B\ell_c^{-d}L^d\log_2{(\hbox{SNR})}$ when $N$ reaches the
{\em saturation} point $\sigma^d_*/\ell^d_c$ in the angular spread 
$\alpha_*$. Here $L^d\ell_c^{-d}=\sigma^d_*\ga^{-d}$ has
the physical meaning of
the {\em angular density} of uncorrelated propagation paths in the medium.

Let us point out several possible  extensions of our results.
First, the case of even broader bandwidth of $0<\lim B\ga\leq 1$
can easily be treated by partitioning the full passband
into many sub-bands with their own $B$ and $\ga$ satisfying
(\ref{uwb}). Since the self-averaging takes place in each sub-band and the whole process is linear, stable super-resolution is valid
in the full passband. Second,  in  the case of a finite time-window, the out-put signals, unlike (\ref{mr}),  involve a coupling of nearby wavenumbers \cite{2f-tire}.
If the time window 
is sufficiently large, $\gg\beta_c^{-1}$, then the coupling takes place only between
wavenumbers of separation much smaller than $\beta_c$
and our result carries over  without major
adjustment. Finally, our results may also be
extended to time-varying channels, prevalent in mobile  wireless
communications, with a low
spread factor $T_c^{-1}\delta_*\ll 1$ where
$T_c$ is the 
coherence time \cite{Paul}. 

\end{document}